\documentclass[12pt,preprint]{aastex}
\usepackage[]{natbib}

\usepackage{times}
\usepackage{graphicx}
\input epsf
\begin{document}



\renewcommand\refname{References and Notes}



\def\plotone#1{\centering \leavevmode
\epsfxsize=\columnwidth \epsfbox{#1}}

\def\wisk#1{\ifmmode{#1}\else{$#1$}\fi}

\def\lt     {\wisk{<}}
\def\gt     {\wisk{>}}
\def\le     {\wisk{_<\atop^=}}
\def\ge     {\wisk{_>\atop^=}}
\def\lsim   {\wisk{_<\atop^{\sim}}}
\def\gsim   {\wisk{_>\atop^{\sim}}}
\def\kms    {\wisk{{\rm ~km~s^{-1}}}}
\def\Lsun   {\wisk{{\rm L_\odot}}}
\def\Zsun   {\wisk{{\rm Z_\odot}}}
\def\Msun   {\wisk{{\rm M_\odot}}}
\def\arcmin  {\wisk{^\prime}}
\def\arcsec  {\wisk{^{\prime\prime}}}
\def\micron  {\wisk{\mu{\rm m}}}
\def\farcs  {\wisk{\atop^{\prime\prime}}}
\def\um     {$\mu$m}
\def\mic     {\mu{\rm m}}
\def\sig    {\wisk{\sigma}}
\def\etal   {{\sl et~al.\ }}
\def\eg     {{\it e.g.\ }}
 \def\ie     {{\it i.e.\ }}
\def\bsl    {\wisk{\backslash}}
\def\by     {\wisk{\times}}
\def\half {\wisk{\frac{1}{2}}}
\def\third {\wisk{\frac{1}{3}}}
\def\nwm2sr {\wisk{\rm nW/m^2/sr\ }}
\def\nw2m4sr {\wisk{\rm nW^2/m^4/sr\ }}

\title{New measurements of the cosmic infrared background fluctuations in
deep Spitzer/IRAC survey data and their cosmological implications}

\author{
A. Kashlinsky\altaffilmark{1,4,6}, R. G. Arendt\altaffilmark{2,4}, M. L. N. Ashby\altaffilmark{5}, G. G. Fazio\altaffilmark{5}, J. Mather\altaffilmark{3,4}, S. H. Moseley\altaffilmark{3,4}
\altaffiltext{1}{SSAI}
\altaffiltext{2}{UMBC}
\altaffiltext{3}{NASA}
\altaffiltext{4}{Observational Cosmology Laboratory, Code 665, Goddard Space Flight Center, Greenbelt MD 20771}
\altaffiltext{5}{Center for Astrophysics, Cambridge, MA 02138}
\altaffiltext{6}{alexander.kashlinsky@nasa.gov}
}





\begin{abstract}
We extend previous measurements of cosmic infrared background (CIB) fluctuations
to $\lsim 1^\circ$ using new data from the Spitzer Extended Deep Survey. Two
fields, with depths of $\simeq 12$ hr/pixel over 3 epochs, are analyzed at 3.6
and 4.5 \um. Maps of the fields were assembled using a self-calibration method
uniquely suitable for probing faint diffuse backgrounds. Resolved sources were
removed from the maps to a magnitude limit of mag$_{AB} \simeq 25$, as indicated
by the level of the remaining shot noise. The maps were then Fourier-transformed
and their power spectra were evaluated. Instrumental noise was estimated from
the time-differenced data, and subtracting this isolates the spatial
fluctuations of the actual sky. The power spectra of the source-subtracted
fields remain identical (within the observational uncertainties) for the three
epochs indicating that zodiacal light contributes negligibly to the
fluctuations. Comparing to 8 \um\ power spectra shows that Galactic cirrus
cannot account for the fluctuations. The signal appears isotropically
distributed on the sky as required for an extragalactic origin. The CIB
fluctuations continue to diverge to $>10$ times those of known galaxy
populations on angular scales out to $\lsim 1^\circ$. The low shot noise levels
remaining in the diffuse maps indicate that the large scale fluctuations arise
from the spatial clustering of faint sources well below the confusion noise. The
spatial spectrum of these fluctuations is in reasonable agreement with an origin
in populations clustered according to the standard cosmological model
($\Lambda$CDM) at epochs coinciding with the first stars era.
\end{abstract}


\section{Introduction}
\label{sec:intro}
The Cosmic Infrared Background (CIB) is the collective radiation emitted throughout cosmic history, including from sources
inaccessible to current telescopic studies (see review by Kashlinsky, 2005).
The latter category includes the very first
luminous objects, currently a subject of intense investigations
and great importance for astronomy and cosmology.
CIB fluctuations can be
more readily discerned than the actual mean level of the background allowing to overcome the significant Galactic and Solar system foregrounds at these wavelengths (Kashlinsky et al. 1996a,b, Kashlinsky \& Odenwald 2000). It is generally believed now as a result of detailed numerical simulations of the formation of first
structures in the standard cosmological model that the first objects to form in the Universe, the so-called
Population III stars, were very massive stars that occupied a brief era at epochs inaccessible to direct telescopic
observations (see review by Bromm \& et al. 2009). Because distant galaxies and pre-galactic
structures are clustered, the CIB has angular fluctuations with a
distinct spectral and spatial signal including a possibly measurable contribution
from the first stars era (Bond et al. 1986, Kashlinsky et al. 2004, Cooray et al. 2004).
The distribution on the sky of the luminous objects to form at early times should be considerably different from the
cosmic pattern seen today, with the differences diverging toward large angular scales and being particularly prominent between 5$^\prime$ to 1$^\circ$.
Although the individual sources at very high $z$ are too faint to observe on their own, fluctuations in the
intensity of the cosmic infrared background (CIB) will reflect the
distribution of those early objects after foreground sources are removed to sufficiently faint levels.

In the course of the prior work we discovered
significant source-subtracted  CIB fluctuations
on scales as large as $\sim 5^\prime$
in deep {\it
Spitzer} IRAC (3.6--8 $\mu$m) data (Kashlinsky et al. 2005, 2007a,b,c - hereafter KAMM1,2,3,4). Extensive details of our analysis and a multitude of tests it was subjected to along with a summary of requirements any reliable CIB data analysis should meet in the presence of significant foreground and instrumental effects is given in Arendt et al. (2010, hereafter - AKMM).
As thoroughly documented in AKMM, we have been very careful
with the processing of the individual IRAC exposures into complete deep mosaicked
images.
AKMM include details with the numerous thorough checks to confirm that we are not being misled by
instrumental artifacts or variations of the zodiacal light. We test that we
see similar background fluctuations in various fields located at different
ecliptic and Galactic latitudes. While all our background fluctuation measurements
have been derived from IRAC data, the recent results of Matsumoto et al. (2011) find similar background fluctuations
using completely independent data from the {\it Akari} satellite, and with performing an
independent analysis. This is the further confirmation that the fluctuations are
not caused by any hidden error in the IRAC data or our analysis.

The residual CIB fluctuations remain after removing galaxies down to very faint
levels and must arise from populations with a
significant clustering component, but only low levels of the shot
noise. As suggested by KAMM1 these CIB fluctuations may originate in early
populations. Further, it was demonstrated by KAMM4 that there are no correlations between the
source-subtracted IRAC maps and the faintest resolved sources observed
with the {\it HST} ACS at optical wavelengths, which likely points
to the high-$z$ origin of the fluctuations, or at least to a very
faint population not yet observed by other means. The high-$z$ interpretation of the detected CIB anisotropies has received strong further confirmation in the recent {\it Akari} data
analysis which, in addition, measured source-subtracted CIB fluctuations at 2.4 \um\ and pointed out that the colors of the fluctuations require them being produced by highly redshifted very luminous sources (Matsumoto et al. 2011). It is inherently important to advance the understanding of the nature and the epochs
of the populations producing these CIB fluctuations with new studies which can be obtained measuring the spatial distribution of the fluctuations at much larger angular scales. This was one of the motivations of the SEDS program (Fazio et al. 2008) and the first results from it are presented below.

This study followed the procedures outlined in AKMM and our previous publications (KAMM1-4) and the paper is organized as follows: Sec. \ref{sec:data} describes the data and its assembly, followed by Sec. \ref{sec:fourier} which lays out the quantities computed from the assembled data. In Sec. \ref{sec:results} we present the results on the spatial spectrum of the source-subtracted fluctuation in the diffuse light remaining after subtracting the instrument noise. Sec. \ref{sec:foregrounds} presents comprehensive tests done to isolate the various non-cosmological contributions to the results and both low- and high-$z$ contributors are discussed in Sec. \ref{sec:interpretation}, where it is shown that the signal cannot be accounted for by the observed populations of known galaxies and requires either unknown new populations at low-$z$ made of low-luminosity systems and distributed very differently from the rest of the present day galaxies, or is dominated by the high-$z$ populations at epochs associated with ``first stars" and clustered according to the established concordance $\Lambda$CDM model.

\section{Data assembly and processing}
\label{sec:data}

The {\it Spitzer Space Telescope} is a 0.85 m diameter telescope launched into an earth-trailing solar orbit in 2003 (Werner et al. 2004, Gehrz et al. 2007). For nearly 6 years, as
it was cooled by liquid He, its three scientific instruments provided imaging and spectroscopy at wavelengths from 3.6 to 160 $\mu$m.
In the time since the He supply was exhausted, {\it Spitzer} has continued to provide 3.6 and 4.5 $\mu$m imaging with its Infrared Array
Camera (IRAC). The IRAC camera has a $5'\times5'$ field of view, and a pixel scale of $1.2''$, which slightly undersampled the
instrument beam size of $\sim2''$ (FWHM) (Fazio et al. 2004a).

The SEDS program is designed to provide deep imaging at 3.6 and 4.5 \um\ over a total area of about 1 square degree, distributed
over 5 well-studied regions (Fazio et al. 2008). The area covered is about ten times greater than previous {\it Spitzer} coverage at comparable depth.
While the main use of the SEDS data sets will be the investigation of the individually detectable and countable galaxies, the remaining
backgrounds in these data are well-suited for CIB studies, by virtue of their angular scale, sensitivity and observing strategies.
The first 2 of the SEDS fields to be completed have been used in this study: the UltraDeep Survey field (UDS; Program ID = 61041) and the Extended Groth Strip (EGS; Program ID = 61042).
The locations, sizes, and depths of these fields are
listed in Table 1.
The fields are located at moderate to high Galactic latitudes to minimize the number of foreground stars
and the brightness of the emission from interstellar medium (cirrus). These fields also lie at relatively high ecliptic latitudes,
which helps minimize the brightness and temporal change in the zodiacal light from interplanetary dust. The observations of each field are
carried out at three different epochs, spaced 6 months apart.

For CIB fluctuations analysis in the assembled maps we have selected regions of approximately uniform exposure in {\it both} Channel 1 and 2. These covered a square region of approximately 21$^\prime$ on the side in the UDS field and a rectangular region of $\simeq 8^\prime\times 1^\circ$ in the EGS field.

The procedure for map assembly is described in our previous papers (KAMM1-4) with an extensive summary including all the tests given in (AKMM). Below we briefly revisit the adopted formalism and its advantages over the standard methods when it comes to removing instrumental artifacts without introducing spurious correlations. This is critically important if one were to robustly measure fluctuations signals as faint as that expected from first stars era, $\delta F \lsim 0.1$ nW/m$^2$/sr at arcminute scales.

The individual 100 sec frame time exposures are processed by the standard IRC calibration pipeline,
and then additionally corrected for the ``column-pulldown effect'' which affects the detector output
in columns that contain very bright stars. The individual frames are then further processed and
combined into mosaicked images (with a $1.2''$ pixel scale) using the self-calibration procedure that we have previously
employed for this work. The 3.6 and 4.5 $\mu$m images are processed independently.
At each wavelength, the frames are also processed in several different groups to provide
multiple images that can be used to assess random and systematic errors. One set of three groups
is obtained by separating the data by each of the three epochs. Comparison of results between
these three sets can provide indications of systematic errors induced by variations in
the zodiacal light over 6 month intervals.
The other set of two groups is obtained by separating the full sequence of frames into the
alternating even and odd frame numbers. Comparison of results from these ``A'' and ``B'' subsets
provides a good diagnostic of the random instrument noise, because the A and B subsets only
differ by a mean interval of $\sim100$s.

We applied the least-squares self-calibration procedure described by
(Fixsen et al. 2000). The approach formalizes the
calibration procedure by describing the data with parameters that
include both the detector characteristics and the true sky
intensity. The derivation of these parameters via a least-squares
algorithm yields an optimal solution for the calibration and the
sky intensity. In this case our chosen model is given by
\begin{equation}
D^i = S^{\alpha} + F^p + F^q
\end{equation}
where $D^i$ represents the raw data from a single pixel of a
single frame, $S^{\alpha}$ is the sky intensity at location
$\alpha$, $F^p$ is the offset for detector pixel $p$,
and $F^q$ is a variable offset for each of the 4 readouts
(alternate vertical columns of the detector) and each frame. This
assumes that the sky intensity ($S^{\alpha}$) and the detector
offsets ($F^p$) are invariant during the
course of the observations. For a data set with fixed frame times
(as our IRAC data), the detector dark current is included in the
$F^p$ term as it is indistinguishable from an offset. For data
sets with multiple frame times, a relatively simple extension of
this data model could be applied. The variable offset $F^q$ can
absorb time--dependent behavior of the detector, but only to the
extent that it can be characterized with a single value per frame, or
in some cases, a single value per readout per frame.

While technically the CIB is the sum of all extragalactic emission,
here we wish to exclude individually detectable objects and analyze the
remaining CIB which is produced by fainter sources.
The mosaicked images were cleaned of resolved sources in two stages.
In the first stage, we construct and subtract an iterative ``Model''
of all the sources in the image. This is done using a variant of
the CLEAN algorithm, in which our iterative loop consists of
(a) identifying the brightest pixel in the image, and (b)
subtracting a scaled version of the instrument PSF (including the
broad low-level wings) to reduce this intensity by a set fraction
(chosen as 50\%). Subtracting only a fraction of a source at each iteration
allows the Model to compensate for sources that are intrinsically
extended in size, and for inaccuracies in the PSF.
Various criteria for selecting the final iteration are given in
AKMM; all lead to very similar results. In order to compare the signal from
different sky locations, the final iteration can be chosen to correspond
to a given shot-noise level.
The second stage involves the construction of masks which are used to set
areas at the locations of bright sources to zero. If the Model worked perfectly, this
would not be necessary. However, small differences between the Model's ideal
PSF and the effective PSF of the mosaicked images lead to relatively
large amplitude residuals at the locations of bright stars. Therefore
we construct a mask which is used to zero all pixels above an effective $4\sigma$
surface brightness. All 8 neighbors of any such pixel are also set to zero. This clipping is done independently of the Model-subtraction and the procedure is iterated until no new pixels above the fixed threshold remain.
The fraction of unmasked pixels is
$\sim 73\%$ which enables robust FFT analysis.

\begin{figure}[h!]
\centering \leavevmode \epsfxsize=1 \columnwidth
\epsfbox{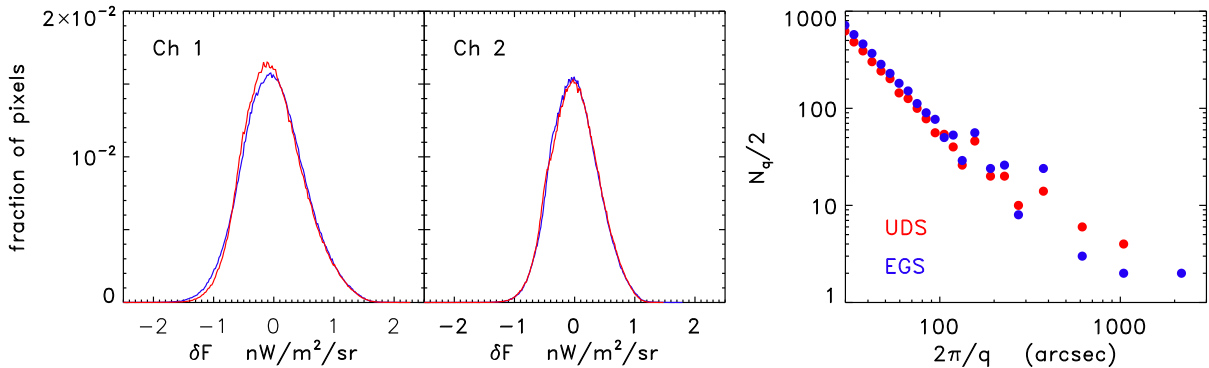} \vspace{0.5cm} \caption[]{\small{Left: Histograms of fluctuations in the EGS (blue) and UDS (red) fields for Ch1 and 2. Right: Number of independent Fourier elements that went into determining the power spectrum for each field and the Fourier binning specified above. } } \label{fig:histo}
\end{figure}

The histogram distributions of the remaining pixel fluctuations after removal of the modeled sources and masking are shown in Fig. \ref{fig:histo}. They are to a good accuracy Gaussian with very small residual skewness, $S\equiv \langle (\delta F)^3/\sigma^3$, parameters ($|S|< 0.05$).

In order to evaluate the random noise level of the maps, alternate
calibrated frames for each dataset were mapped into separate ``A''
and ``B'' mosaics. The difference of the A and B mosaics should
eliminate true celestial sources and stable instrumental effects
and reflect only the random noise of the observations. The power
spectrum of the instrument noise is then subtracted from the power
spectrum of the assembled maps to give the measured signal of the
remaining source-subtracted CIB fluctuations in the maps.

\section{Fourier analysis of the assembled data}
\label{sec:fourier}

\subsection{Definitions}

The maps under study are clipped and masked of the
resolved sources, yielding the fluctuation field, $\delta
F(\vec{x})$. Its Fourier transform,
$\Delta(\vec{q})= \int \delta F(\vec{x}) \exp(-i\vec{x}\cdot\vec{q}) d^2x$ is
calculated using the FFT. The power spectrum is $P_2(q)=\langle |
\Delta(\vec{q})|^2\rangle$, with the average taken over
all the independent Fourier elements corresponding to the given $q$. A
typical rms flux fluctuation is $\sqrt{q^2P_2(q)/2\pi}$ on the
angular scale of wavelength $2 \pi/q$. The correlation function, $C(\theta) = \langle \delta F(\vec{x})\cdot \delta F(\vec{x}+ \vec{\theta})\rangle$, is uniquely related to $P_2(q)$ via Fourier
transformation. If the fraction of masked pixels in the maps is too high (e.g. $>40\%$ for IRAC maps as detailed in KAMM1-4/AKMM), the large-scale map properties
cannot be computed using the Fourier transform and instead the maps must be analyzed by
direct calculation of $C(\theta)$, which is immune to mask effects (Kashlinsky 2007).

Several quantities are used in computations below. The power spectrum in a single band $n$ is defined as $P_n(q)$. It is computed after averaging all independent Fourier elements which lie inside the radial interval $[q, q+dq]$. Since the flux is a real quantity, only one half of the Fourier plane is independent, so that at each $q$ there are $N_q/2$ independent measurements of $\Delta$ out of a full ring with $N_q$ data.  Because we will compare two fields of different configurations, the Fourier space of each was binned at the same grid of $q$, the mean power was then evaluated and the relative (Poissonian) errors on that determination were computed as $\sqrt{\half N_q}$.

After demonstrating that different fields have statistically similar power spectrum, the overall power at each $q$ was averaged over the different fields with each field weighted by the relative statistical uncertainty defined by its configuration (see below), i.e. $P(q)=\sum_i P_i N_q^i/\sum_i N_q^i$. This is equivalent to simply averaging over all $|\Delta(q)|^2$ available from each field $i$.

We characterize the similarity (or not) of the fluctuations measured in different channels, or at different Epochs of observations. A quantity of further interest in this context is the cross-power, which is the Fourier transform of the cross-correlation function $C_{mn} (\theta) =\langle \delta F_m(\vec{x})\cdot \delta F_n(\vec{x}+ \vec{\theta})\rangle$. The cross-power spectrum is then given by $P_{mn} (q) = \langle \Delta_m(q) \Delta^*_n(q)\rangle = {\cal R}_m(q) {\cal R}_n(q) + {\cal I}_m(q) {\cal I}_n(q)$ with ${\cal R, I}$ standing for the real, imaginary parts. Note the cross-power of a real quantity, such as the flux fluctuation, is always real, but unlike the single (auto-) power spectrum the cross-power can be both positive and negative. The presence (or not) of the same population at two channels, $(m,n)$, can then be probed via {\it coherence} defined as ${\cal C}_{nm} \equiv \frac{P_{nm}(q)*P_{nm}(q)}{[P_n(q)P_m(q)]}$. If ${\cal C}_{mn}\simeq 1$ then at the two channels the same populations produce the diffuse signal and vice versa.

\subsection{Power spectra of the assembled fields}

The clipped and cleaned maps were Fourier transformed and power spectra evaluated. Visual and numerical inspection of the maps did not show any significant presence of artifacts, stray light, muxbleed etc. In order to average over the different fields the Fourier space was binned at the same set of central wavenumbers, $q$, in both configurations. The values of $q$ chosen for the final $P(q)$ evaluations are shown in the right panel of Fig. \ref{fig:histo} which displays the number of independent Fourier elements which is available for power spectrum evaluations. Since the relative error on the determined $P(q)$ is $[\half N_q]^{-\half}$, the power spectrum is not determined highly accurately at the outer 2 bins of the EGS field; however, when combined with the independent determination in the UDS field, the accuracy of the averaged $P(q)$ is acceptable ($\lsim 30-40\%$) at scales $\leq 1,000^{\prime\prime}$. Only the largest scale of $\simeq 1^\circ$ probed with the EGS field alone has $\half N_q=2$ and is not measured well.

The instrument noise was evaluated from the $\half(A-B)$ maps, where $A$ and $B$ correspond to the sequences of alternating odd/even AORs.


\begin{figure}[h!]
\centering \leavevmode \epsfxsize=0.8 \columnwidth
\epsfbox{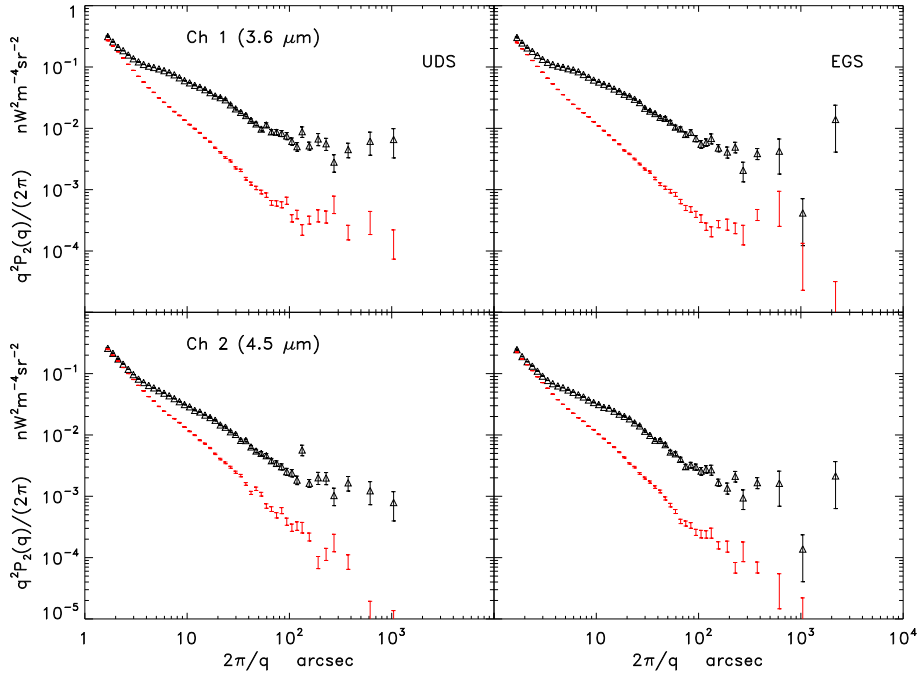} \vspace{0.5cm} \caption[]{\small{Fluctuations from the $(A+B)$ (black triangles) and $\half(A-B)$ maps (red error bars).} } \label{fig:power}
\end{figure}
Figure \ref{fig:power} shows the total power remaining in the source-subtracted maps and the noise. At small scales (within the beam at $\sim 3^\prime-5^\prime$) the noise contributes significantly, as it should, but there is a clear excess in the fluctuation power at larger scales with the noise becoming progressively smaller with its contribution reducing to negligible at scales greater than $\sim 10^\prime-20^\prime$.

\section{Diffuse source-subtracted fluctuations}
\label{sec:results}

\begin{figure}[h]
\centering \leavevmode \epsfxsize=1 \columnwidth
\epsfbox{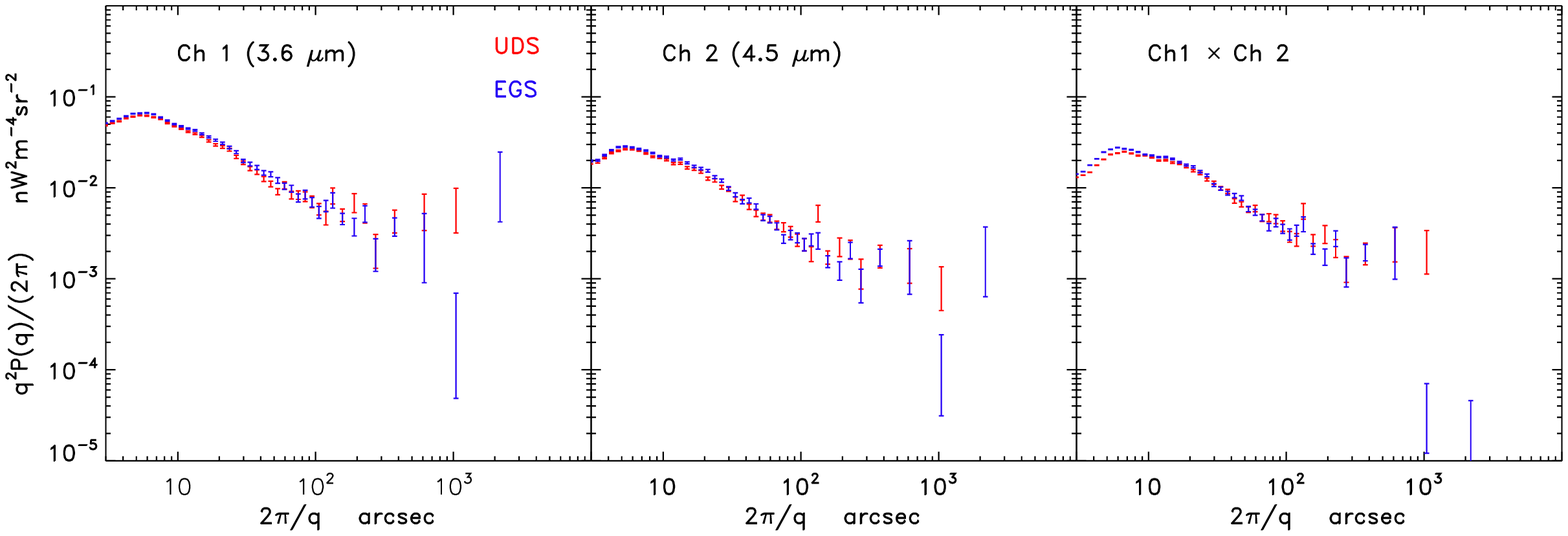} \vspace{0.5cm} \caption[]{ \small{The spatial spectra of mean squared fluctuations for each of the fields in this study. Red/blue error bars correspond to UDS/EGS fields. We verify that after subtraction of the Model, the masked images exhibit no
correlations with the removed Model.
Histograms of the unmasked pixels intensities are approximately Gaussian,
and are shown in Fig. \ref{fig:histo}.
Power spectra $P(q)$ of the CIB are calculated as the amplitudes of the Fourier transforms
of the masked images. The power spectra are average in fixed intervals of
angular frequency $q$ to reduce uncertainties at the largest angular scales, and
to enable direct comparison of power spectra between the two fields
with different sizes and shapes. The relative uncertainties resulting from sample (cosmic) variance
were evaluated as $\sqrt{N_q/2}$ where $N_q$ is the number of Fourier elements averaged
with each interval in $q$ (Fig. \ref{fig:histo}).} } \label{fig:cib_fields}
\end{figure}

The resulting power spectra of the residual CIB after the above subtraction and masking of resolved sources is
shown in Fig. \ref{fig:cib_fields}. Here we have subtracted the instrumental noise power by deducting the
power spectra of the differences of the A and B mosaics [\half(A-B)], which are displayed in Fig. \ref{fig:power}. The instrumental noise power is comparable to the
the CIB power at scales $<3''$, but is much smaller than the power at larger angular scales.
At both 3.6 and 4.5 $\mu$m, the power spectra are in good agreement for EGS and UDS fields, except perhaps
at the largest angular scale ($\geq 1000''$) where the small number of effective measurements leads to
large uncertainties on the power in each individual field.


In order to check for the presence of the same source-subtracted CIB at both 3.6 and 4.5 $\mu$m, we have computed the
cross-correlation power spectrum between the two channels. The right panel of the figure shows the
resultant cross-correlation power spectra. Their similarity to the full power spectra confirm that the fluctuations are
correlated in wavelength for both fields at all angular scales. Similar amplitudes would not be expected if the signal was dominated by noise,
or by instrumental artifacts that occur independently at each wavelength. The correlation of many instrumental effects are
mitigated by the fact that IRAC's 3.6 and 4.5 $\mu$m channels use separate optical systems and detectors to simultaneously
observe fields separated by $\sim6'$.


\section{Non-cosmological contributions}
\label{sec:foregrounds}

The results in Fig. \ref{fig:cib_fields} are in excellent agreement with our earlier measurements in five additional fields (KAMM1-2) as shown in Appendix A, but are now measured to significantly more accurately and extending to much greater angular scales. The isotropy of the signal
(i.e. the consistency of results in different fields)
by itself suggests a cosmological origin of the fluctuations. Nevertheless, we describe below a multitude of additional tests that we have performed in order to verify that non-cosmological sources cannot explain the measured signal.

\subsection{Cross-correlating different epochs of observation}

The data were analyzed separately after combining the AORs in each of the 3 epochs of observations, E1-E3.
Epochs E1 and E2 (and E2 and E3)
are separated by $\sim 6$ months and the detector orientation changes by $\sim 180^\circ$ between each pair of epochs. If the signal arises from the detector, rather than the sky, the fluctuations in each of the epochs should not correlate. Additionally, if zodiacal light fluctuations contribute significantly to the measured signal, the measured fluctuations should differ appreciably between the epochs.

\begin{figure}[h!]
\centering \leavevmode \epsfxsize=1 \columnwidth
\epsfbox{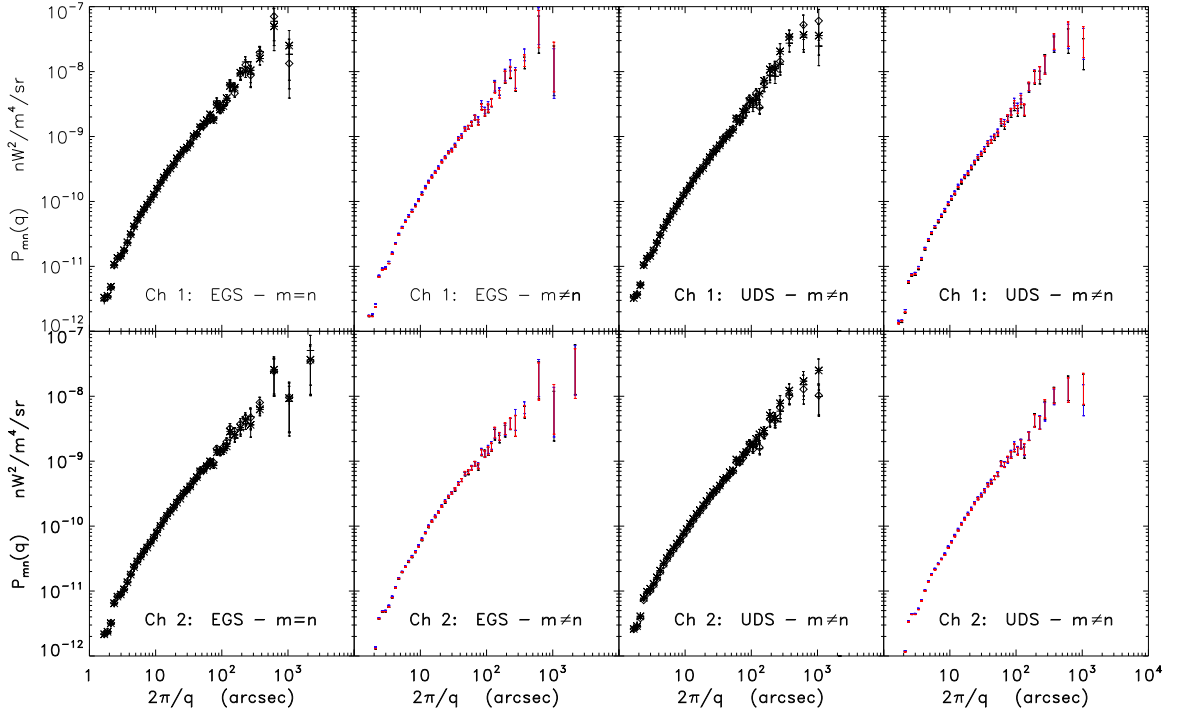} \vspace{0.5cm} \caption[]{\small{Auto ($m=n$) and cross-correlation ($m\neq n$) power spectra between three epochs for EGS and UDS field at the two IRAC channels. Black, blue and red symbols correspond to $E1\times E2, E1\times E3, E2\times E3$ respectively. In the auto-correlation correlation panels the plus signs, asterisks and diamonds correspond to E1, E2, E3 respectively. } } \label{fig:cross}
\end{figure}

The random instrument noise is negligible at scales $\gsim 10^{\prime\prime}$. We cross-correlated the power spectra at epochs separated by 6 months when the detector orientation is rotated by 180$^\circ$. Instrument artifacts
that remain in a fixed geometry with respect to the detector, such as stray light and ghost images, are thus projected on to the
field in very different patterns at the different epochs. The cross-correlations power spectra are shown in
Fig. \ref{fig:cross} and demonstrate that the signal comes from the sky rather than the detector.

\clearpage
\subsection{Instrumental Stray Light}

\begin{figure}[h!]
\centering \leavevmode \epsfxsize=1 \columnwidth
\epsfbox{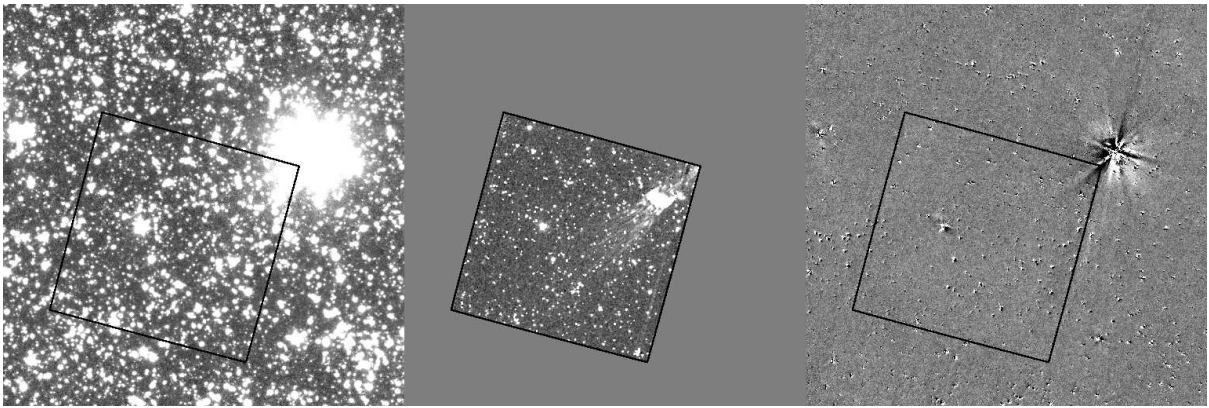} \vspace{0.5cm} \caption[]{\small{(left) A $10'\times10$' section of the 3.6 $\mu$m mosaic of
the UDS field [range = (-0.01,0.01) nW m$^{-2}$ sr$^{-1}$]. (middle) A single 3.5 $\mu$m frame that shows very strong stray
light from the bright star just outside the upper right corner of the frame. This image is displayed on a $10\times$
wider range [(-40,40) nW m$^{-2}$ sr$^{-1}$] to better show the typical structure of bright stray light.
(right) A $10'\times10$' section of the difference of mosaics that were made using (a) only data from the upper
half of the detector, and (b) only data from the lower half of the detector. The range here is the same as in the
left hand panel [(-4,4) nW m$^{-2}$ sr$^{-1}$]. No evident stray light artifacts remain.
The brighter sources do exhibit small changes in the details
of the PSF between the top and bottom halves of the array. These regions are masked in out analysis.}} \label{fig:stray}
\end{figure}

One of the potential difficulties in measurements of backgrounds and their structure is stray light. Stray light will cause
an apparent nonuniform increase in the background across the detector. The intensity and the pattern of the stray light
depends on the distribution of sources {\it outside} of the field of view being observed. In the case of the IRAC 3.6 and 4.5 $\mu$m
detectors, sources that lie $\lesssim1'$ from the top edge of the detector can scatter light back onto the detector \citep{hora}. The geometry
of the scattering leads to two concentrated regions of stray light in the upper left and right portions of the
detector\footnote{http://irsa.ipac.caltech.edu/data/SPITZER/docs/irac/iracinstrumenthandbook/}. Uniform diffuse
illumination, such as the zodiacal light, leads to a fixed pattern on the detector. An individual bright source leaves a more concentrated
and irregular patch of stray light, but at a location that is well defined given the known source position.
An example of this is shown in the left and center panels of Fig. \ref{fig:stray}.
Thus individual IRAC exposures have been
masked\footnote{http://irsa.ipac.caltech.edu/data/SPITZER/docs/dataanalysistools/tools/contributed/irac/straylight/}
to exclude regions that are potentially affected by the stray light of stars identified by brightness and color thresholds
from the 2MASS catalog \citep{2MASS}.
However, there remains some question of whether or
not the stray light from sources at fainter levels might yield some structure in the observed background.

\begin{figure}[h!]
\centering \leavevmode \epsfxsize=1.0 \columnwidth
\epsfbox{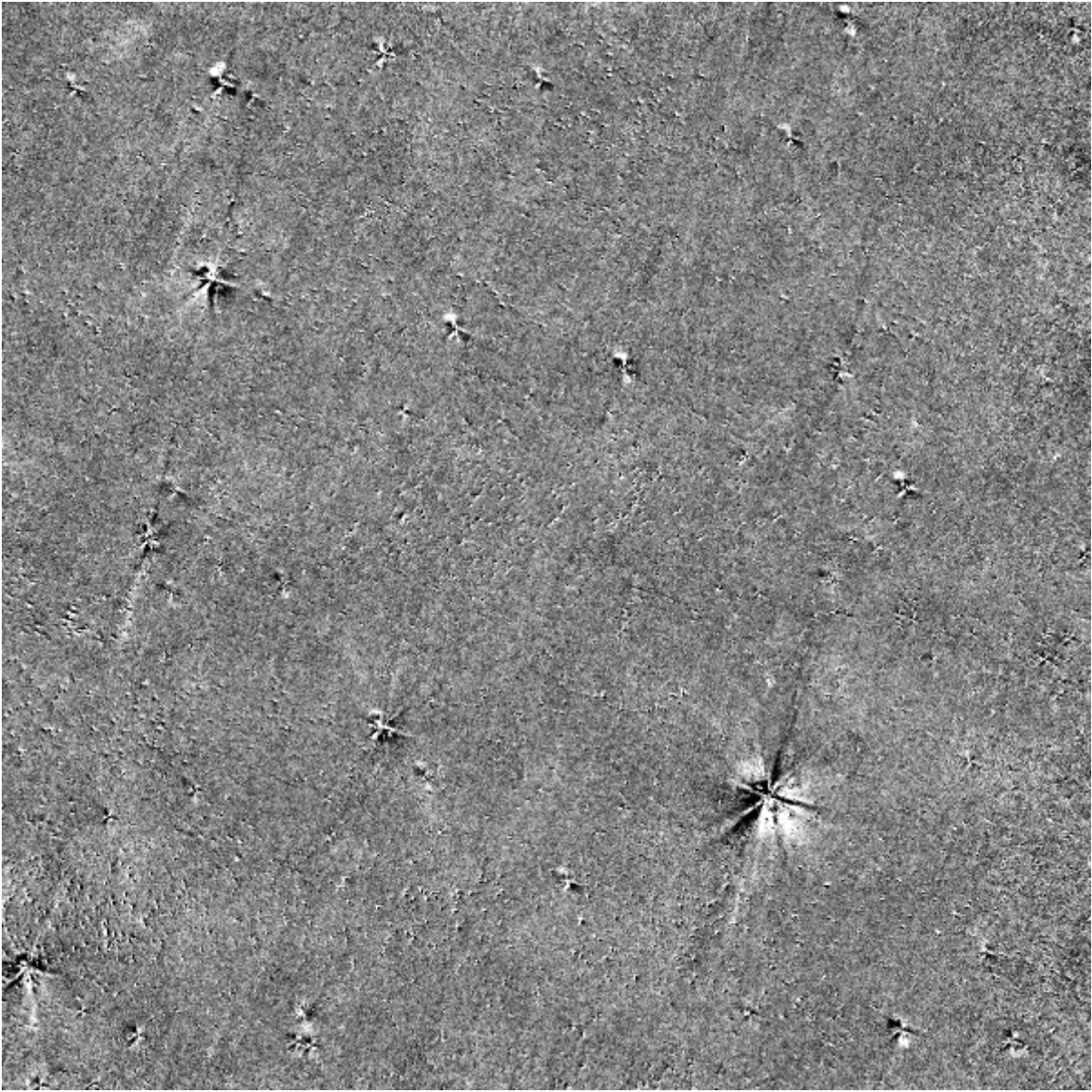} \vspace{0.5cm} \caption[]{\small{A $24'\times24$' section of the difference
of mosaics that were made using (a) only data from the upper
half of the detector, and (b) only data from the lower half of the detector. The range here is the same as in
Fig. \ref{fig:stray} [(-4,4) nW m$^{-2}$ sr$^{-1}$]. If stray light from fainter sources and the background
were present, it would appear as 5 horizontal bands matching the layout of the coverage of the field.}} \label{fig:stray2}
\end{figure}

To address this question, we created new versions of the 3.6 $\mu$m mosaic of the UDS field from two subsets for the data.
The first version is made only
with data from the upper half of detector, which can be affected by stray light if the data are
insufficiently masked. The second version is made from only from data in the lower half of the
detector which is largely unperturbed by stray light. The difference between these two mosaics
would reveal if there were systematic differences between the upper and lower halves of the detector, whether due to
stray light or any other instrumental effect. The difference image
reveals no residual stray light near bright stars [Fig. \ref{fig:stray} (right)] and
obvious large scale structures [Fig. \ref{fig:stray2}] that could arise from the integrated stray light of the many
faint sources.
Although small scale changes in the instrumental PSF between upper and lower halves of the detector are evident
at the locations of bright sources. The power spectrum of the difference image is
very similar to the power spectrum of the \half(A-B) image used to assess the instrumental noise (Fig. \ref{fig:power}).
Therefore we have verified that the observed large-scale fluctuations are not an artifact related to instrumental stray light.

\clearpage

\subsection{Zodiacal light}

Zodiacal light is the brightest source of foreground emissions at near-IR (Kelsall et al. 1998), but appears to be extremely smooth with relative fluctuations below 0.2\% (Abraham et al. 1997). More recently, Pyo et al. (2012)
used mid-IR (7-24 $\mu$m) data {\it Akari} data to establish upper limits of $\sim0.02\%$ at the North Ecliptic Pole.
Expected levels for zodiacal light fluctuations in our measurements have been discussed in the context of the Spitzer-based measurements (KAMM1,2 and AKMM) and also in Matsumoto et al. (2011) for the {\it Akari}-based results.
Because the {\it Spitzer} orbits with the zodiacal cloud, the zodiacal light intensity and the direction of its gradient will vary
with a roughly annual period in each field. The comparison of power spectra at, and cross-correlation power spectra between, the different
epochs also serves as an empirical constraint on the influence of the zodiacal light.

The fields in this study lie at high Ecliptic latitude well outside the Ecliptic plane where zodi is at its highest. In earlier studies it was already estimated by us to be much smaller than the detected fluctuations. The gradient due to zodiacal light would vary as the telescope moves around the Sun and hence zodiacal light would contribute different levels at different epochs.
Then any change in the actual sky brightness or power spectra between three Epochs of observations separated by $\sim 6$ months can be attributed to changes in the zodiacal light.

Fig. \ref{fig:cross} shows the cross-power spectra between the epochs and the signal in each of the epochs for both fields. The figure shows that the signal remains the same in each of the epochs and correlates remarkably well between them. This argues against substantial contribution from either the zodiacal light or detector systematics.

To sum up this part of discussion we find no significant differences between epochs,
demonstrating that zodiacal light is not a significant contributor to the observed spatial fluctuations.
Furthermore, the relatively
constant intensity of the fluctuations from 3.6 to 4.5 $\mu$m is at odds with the measured zodiacal light spectrum
which rises by a factor of ~2.5 from 3.6 to 4.5 $\mu$m.
Fig. \ref{fig:cross} shows that the three separate Epochs of data produce the same fluctuation signal and they all correlate very well with each other. This directly confirms that the contribution of the time-varying component to the measured fluctuations at 3.6 and 4.5 \um\ is small and zodiacal fluctuations cannot account for the measurement.

\subsection{Galactic cirrus}

Galactic cirrus is another source of potential confusion in this measurement. However, its flux levels depend on
the line of sight column density of the Galactic ISM
and lead to variations by factor of a few ($\sim 3-4$) among all the seven fields probed in our studies (KAMM1, KAMM2 and SEDS). On the other hand, the measured fluctuation signal at 3.6 and 4.5 \um\ is the same, within the statistical uncertainties, at all locations. This by itself argues against a significant contribution of the Galactic cirrus emission.
%
%


Contributions of Galactic cirrus to the source-subtracted fluctuations at the near-IR wavelengths have been addressed earlier in our Spitzer analysis (KAMM1-2 and AKMM) and in the {\it Akari}-based study (Matsumoto et al. 2011). All of the studies reached the conclusion that Galactic cirrus contributes well below the fluctuation levels measured there. In addition, in the {\it Akari}-based analysis it was demonstrated that the near-IR source-subtracted maps do not correlate with the same maps observed at 100 \um, which is the most accurate probe of cirrus emission (Arendt et al. 1998, Schlegel et al. 1998). In order to further verify that the Galactic cirrus emission cannot contribute appreciably to the signal measured at 3.6 and 4.5 \um\ we proceed as outlined in KAMM1,2 and AKMM with some extra steps. We measure identical fluctuations in both fields in both IRAC bands despite the fact that the mean cirrus flux varies by factors of a few from the UDS to the EGS. The level of the fluctuations also agrees well with our earlier measurements at five additional locations out to angular scales of $\sim 5^\prime$ (KAMM1,KAMM2).

As these high latitude deep survey fields are chosen in part based on having low ISM column densities, it is no surprise that we
see no direct evidence of cirrus in these fields. Cirrus emission at 3.6 and 4.5 $\mu$m is largely continuum emission from
stochastically (transiently) heated small dust grains and polycyclic aromatic hydrocarbons (PAHs). An empirical measurement of
cirrus emission needs to be made at longer wavelengths, such as 100 $\mu$m where the emission arises from the bulk of the
ISM dust (larger grains at cooler temperatures), or at 8 $\mu$m where the emission is dominated by strong PAH emission
bands that are closely related to the 3.6 and 4.5 $\mu$m emission.

\begin{figure}[t!]
\centering \leavevmode \epsfxsize=0.9 \columnwidth
\epsfbox{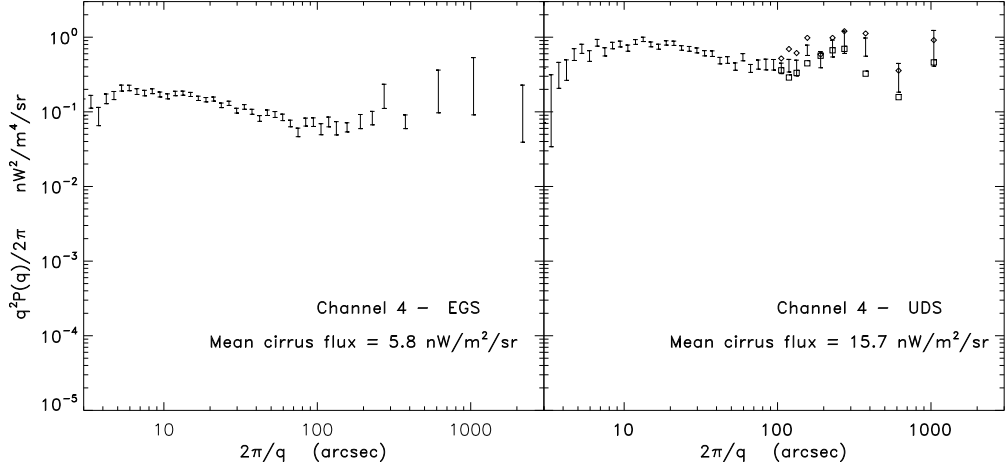} \vspace{0.5cm} \caption[]{\small{Source subtracted CIB fluctuations at 8 \um\ for EGS and UDS fields. In the UDS field at 8\um\ the maps contain a notable large-scale artifact which translates into a diagonal stripe in Fourier space. To probe the sensitivity of the resultant power spectrum to this feature we have computed the power spectrum independently from the two quadrants in the Fourier space: $(q_x<0, q_y>0)$ (containing the stripe feature) and $(q_x>0, q_y>0)$. The power spectra from these are shown with open diamonds and squares respectively and can be seen to be within the statistical uncertainties of the overall spectrum.} } \label{fig:cirrus1}
\end{figure}

As in our prior studies, in order to evaluate cirrus contributions we have also examined IRAC
8 \um\ images generated from earlier observing programs carried out while {\it Spitzer} was still operating cryogenically. 
These data are taken with the same IRAC instrument, and
have similar angular resolution to the 3.6 and 4.5 $\mu$m data. The 8 $\mu$m data are somewhat shallower than the
SEDS observations, as the IRAC 8 $\mu$m channel was only in operation during earlier cryogenic observations of the EGS and UDS
fields. The UDS field was covered to an average depth of 0.67 hr under Spitzer program ID = 40021, and the EGS filed was coved to a mean depth of 1.44 hr under Spitzer program ID = 8.
The 8 \um\ UDS and EGS images were then masked with the masks from 3.6 and 4.5 $\mu$m and the power spectra computed.
They are shown in Fig. \ref{fig:cirrus1}.
Their amplitudes at large scales are roughly proportional to the mean cirrus
intensity predicted (by extrapolation of 100 $\mu$m observations) for these fields.
However, this extrapolation is not precise because of intrinsic variation in the energy spectra of the cirrus.
Therefore, these power spectra should be viewed as setting a strict upper limit on the possible cirrus
contribution at 8 $\mu$m.
The 8 \um\ fluctuations are also expected to contain contributions from the sources producing the measured signal at 3.6 and 4.5 \um\, as well as that from other populations, so assuming that {\it all} of the 8 \um\ fluctuations signal is produced by cirrus gives the most conservative assumption about its contribution.

\begin{figure}[h!]
\centering \leavevmode \epsfxsize=0.9 \columnwidth
\epsfbox{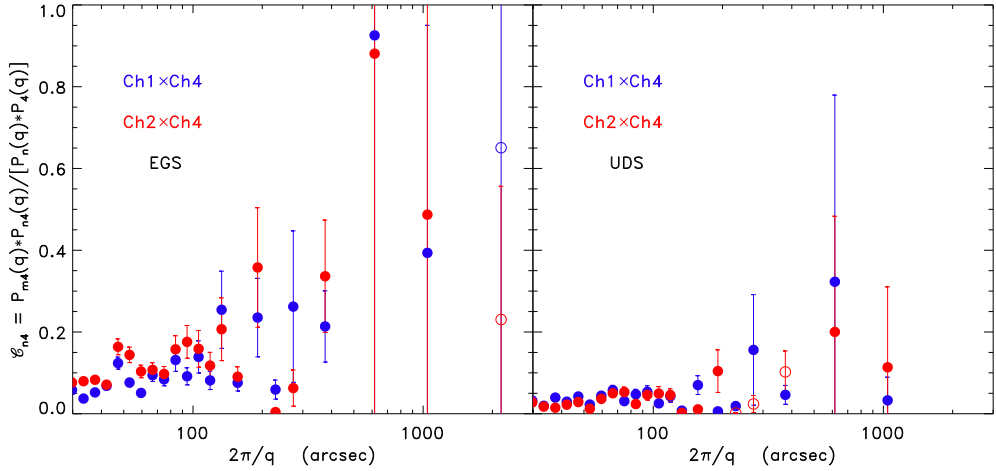} \vspace{0.5cm} \caption[]{\small{Coherence, ${\cal C}_{n4} \equiv \frac{P_{n4}(q)*P_{n4}(q)}{P_n(q)P_4(q)}$,  for 1-4 (blue) and 2-4 (red) for EGS (left) and UDS (right)  fields. Open symbols correspond to scales where $P_{n4}<0$. The relative statistical uncertainties on the coherence resulting from cosmic variance are given by $\sqrt{12/N_q}$ which is shown with the error bars; this is valid at small ${\cal C}$ when all the power spectra can be assumed independent. With $N_q$ plotted in Fig. \ref{fig:histo} this uncertainty is of order 100\% for the last three points in the EGS field.
} } \label{fig:cirrus2}
\end{figure}

In order to estimate directly whether cirrus provides a significant contaminant of the measured fluctuations, we assume that {\it all} of the measured 8 \um\ diffuse emission
fluctuation on large angular scales (where shot-noise is negligible) is generated by cirrus and then computed the cross-power spectrum expressing its value in terms of the {\it coherence} between channel $n$ and Channel 4 at 8\um, ${\cal C}_{n4}\equiv \frac{P_{n4}P_{n4}}{P_nP_4}$. If cirrus emission contributed all of the Channel 4 signal and provided a significant contribution to the measurements at 3.6 and 4.5 \um\, the coherence would have to be ${\cal C}_{n4}\simeq 1$. However, the results of the computation are shown in Fig. \ref{fig:cirrus2} and indicate that, even if dominant at 8 \um, Galactic cirrus contributes negligibly to the measured signal at 3.6 and 4.5 \um. The coherence of the EGS field is largely uncertain at the largest scale 3-4 data points, where the cross-power $P_{n4}$ is sometimes {\it negative}, but as mentioned above, the levels of cirrus fluctuations there are expected to be at a level similar to that at smaller scales\footnote{Since the cirrus power spectrum has power slope of $n\sim -2$ (e.g. Kashlinsky \& Odenwald 2000), one expects the level of cirrus flux fluctuations at $\lsim 1^\circ$ to be similar to that at the smaller scales.}
 and also the UDS field shows that cirrus contributes negligibly at the largest scales probed here.

The energy spectrum of the cirrus drops by a factor of $\gsim 10$ from 8 to 4.5 and 3.6 $\mu$m. Therefore, even if the amplitude of the
large-scale power spectra at 8 $\mu$m is dominated by cirrus, the amplitudes at 3.6 and 4.5 $\mu$m will be at least $\sim 100$
times lower. At such levels, it remains possible that the cirrus could account for the large scale power spectra. Yet this is an unlikely
scenario because calculation of the {\it coherence} expressed via cross-power spectra, $P_{14}, P_{24}$ (Fig. \ref{fig:cirrus2}) shows
a relatively weak correlation between the 8 $\mu$m and shorter wavelength emission. This conclusion is in agreement with that of the Akari analysis (Matsumoto et al. 2011) who demonstrate that the near-IR fluctuation do not correlate with the 100 \um\ data for the same location implying a negligible contribution of cirrus to the former.


\section{Cosmological implications}
\label{sec:interpretation}

Having established the likely cosmological origin of the measured fluctuations, we now turn to interpreting these results.

There is a statistically significant signal on the remaining source-subtracted CIB fluctuations in the Spitzer data. It now extends to angular scales $\lsim 1^\circ$ and, as demonstrated in Appendix A is consistent with our earlier measurements. KAMM1 have proposed that these CIB fluctuations originate at early epochs and KAMM3 have discussed the properties of the populations needed to reproduce the measured signal. On the other hand, the high-$z$ interpretation has been challenged by Cooray et al. (2007) who suggested that faint ACS galaxies at intermediate $z$ are responsible for the measured  signal at 3.6 \um\ (no results for the 4.5 \um\ band IRAC fluctuations are presented there) because they claim a drop in power at 3.6 \um\ when ACS-identified galaxies are blanked out\footnote{Two problems with the Cooray et al. analysis have been discussed as follows: 1) the power spectrum there is computed when over 70-80\% of the field is lost to the mask which makes FFT-based analysis suspect. Kashlinsky (2007) shows that - when the data from Cooray  et al. are analyzed via the correlation function, which is immune to masking - no such drop occurs. 2) Additionally, AKMM in Fig. 5 demonstrate significant inadequacies of the image assembly utilized in Cooray et al.} and by Thompson et al. (2007b) who proposed that the energy spectrum of the fluctuations is consistent with colors of stellar populations at $z\le8$. AKMM (see Secs. 4.2, 6.3.3 and Figs. 5, 38 there) have discussed both of these proposals (Cooray et al. 2007, Thompson et al. 2007b) in detail and showed that high-$z$ interpretation remains the best description of all the previous data.

\begin{figure}[h!]
\centering \leavevmode \epsfxsize=1. \columnwidth
\epsfbox{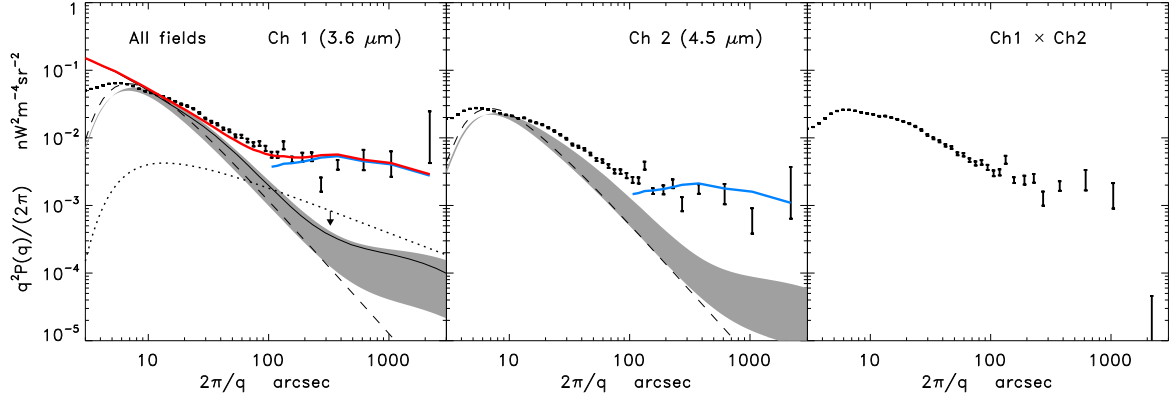} \vspace{0.5cm} \caption[]{ \small{Field-averaged CIB fluctuations at 3.6 (left), 4.5 \um\ (middle)  and the cross-correlation power spectrum.
Black solid line is the contribution of the remaining known galaxies from Sullivan et al. (2007) who state that normal galaxies at Vega magnitudes from 22.5 -- 26 can fit the observed large scale fluctuations. [That claim, which also appears in the {\it comments} to their arXiv:astro-ph/0609451 posting, is contradicted by their own Figure 8 which shows the clear deficit compared to the KAMM1 measurements]. Because Sullivan et al. (2007) present their results only for 3.6 \um\ sources, their model is displayed only in the left panel. Shaded areas show the residual fluctuations from Helgason et al. (2012), after reconstructing the near-IR CIB fluctuations of known galaxies at both 3.6 and 4.5 \um\ from a zoo of multiband galaxy luminosity function (LF) data. The shaded regions correspond to the high- and low faint end of the LF data. At 3.6 \um\ they are consistent with the black solid line although Helgason et al find, on average, slightly lower levels compared to Sullivan et al. (2007). The dashed line shows the shot-noise contribution: at 3.6 \um\ the regression leads to $P_{\rm SN} = 57.5$ nJy$\cdot$nW/m$^2$/sr (or $4.8 \times 10^{-11}$ nW$^2$/m$^4$/sr) and at 4.5 \um\ the shot noise levels are $P_{\rm SN} = 31.5$ nJy$\cdot$nW/m$^2$/sr (or $2.2 \times 10^{-11}$ nW$^2$/m$^4$/sr). Since the shot noise can be expressed as $P_{\rm SN}\sim S F_{\rm CIB}(>m_0)$ it is presented in both sets of units. Blue solid line corresponds to the high-$z$ $\Lambda$CDM (toy)-model processed through the mask of each field and then averaged as described in the Sec. \ref{sec:masking}. It leads to the fiducial amplitude at 5$^\prime$ of $A_{5^\prime}=0.07(0.05)$ nW/m$^2$/sr at 3.6(4.5) \um. The thick solid red line shows the sum of the three components. } } \label{fig:cib_final}
\end{figure}

Fig. \ref{fig:cib_fields} shows that, within the statistical uncertainties, the fields in this study have the same power spectrum of the source-subtracted fluctuations. Therefore, we averaged the two sets of results to obtain an overall power spectrum describing the CIB. The averages were weighted with the number of Fourier elements in each field as described above. The results of the CIB fluctuations after averaging over the individual fields are shown in Fig. \ref{fig:cib_final}. The figure indicates the presence of significant large-scale fluctuations remaining after removing/subtracting the resolved sources.

The measured fluctuation spectrum is made of two components: small scales, $\lsim 10^{\prime\prime}$, are
dominated by the shot-noise produced by the discreteness of the remaining sources. The isotropy of the measured signal, which is further demonstrated in Appendix A for five additional fields from our prior measurements, is consistent with it being of cosmological origin. At the same shot-noise level, the measured signal is in excellent agreement with our measurements at five other sky locations (KAMM1-2) at smaller angular scales
($\lsim 300^{\prime\prime}$).

This section discusses the constraints on the populations producing both the shot-noise and clustering components. It is organized as follows: we first address the limitations on the fluxes of the individual sources producing the large-scale clustering component which stem from the measured shot-noise levels and galaxy counts. Next we revisit, in light of the new results, the existing estimates of the levels of the clustering component of the CIB fluctuations from the remaining known galaxy populations and emphasize, again, that the known galaxy populations do not account for the large-scale clustering component, both its amplitude and spatial dependence. Finally, we discuss the limitations on the nature and epochs of the new populations implied by both the measured shot-noise levels and the clustering component of the source-subtracted CIB fluctuations.


\subsection{Galaxy counts and shot noise levels}

On scales greater than $\sim 20^{\prime\prime}$
the fluctuation spectrum is dominated by a component due to clustering of the sources producing these CIB fluctuations. On scales from
$\sim100^{\prime\prime}$ to $\sim 1000^{\prime\prime}$ the amplitude of this component remains roughly constant at
$(\delta F_{\rm clus})^2=q^2P(q)/(2\pi) \sim 5\times10^{-3}$ nW$^2$/m$^4$/sr$^2$ at 3.6 \um\ and
$\sim 2\times10^{-3}$ nW$^2$/m$^4$/sr$^2$ at 4.5 \um.
The shot noise of the sources that produce these large scale fluctuations cannot exceed the observed shot noise component
that dominates the fluctuations at scales $\lesssim10''$.
The shot noise level is measured to be $P_{\rm SN} \sim 60, 30$ nJy$\cdot$nW/m$^2$/sr at 3.6, 4.5 \um\ respectively.
These levels can be compared to the shot noise derived from the observed galaxy counts in the fields to set lower limits
on the magnitudes (upper limits on the flux densities) of the sources that remain unmodeled and unmasked in the background.

Galaxy counts were derived independently from the reduction for the background fluctuation analysis.
The {\sl Spitzer}/IRAC SEDS data for the EGS and UDS fields were reduced
independently to mosaic form using a combination of facility
pipeline processing tools and IRACproc (Schuster et al. 2006).  The resulting mosaics were constructed with 0.6$^{\prime\prime}$ pixels
using temporal outlier rejection to eliminate cosmic rays and
instrumental artifacts. We performed source extraction using SExtractor (Betin \& Arnouts 1996)
in both fields and both bands -- a total of four mosaics.  This was done
in dual-image mode, using the 3.6\,$\mu$m mosaic as the detection image
and then carrying out photometry only at the positions of detected
sources in the mosaics.  To account for source confusion we inserted
10,000 randomly-placed point sources in the mosaics and then attempted
to recover them using SExtractor with identical parameter settings as
were used for the original source detection.  In this way the source
counts -- which are dominated by galaxies below 16 Vega mag -- could be
corrected for completeness.  For more details see Ashby \etal (2012), in
preparation.

\begin{figure}[h!]
\centering \leavevmode \epsfxsize=0.7 \columnwidth
\epsfbox{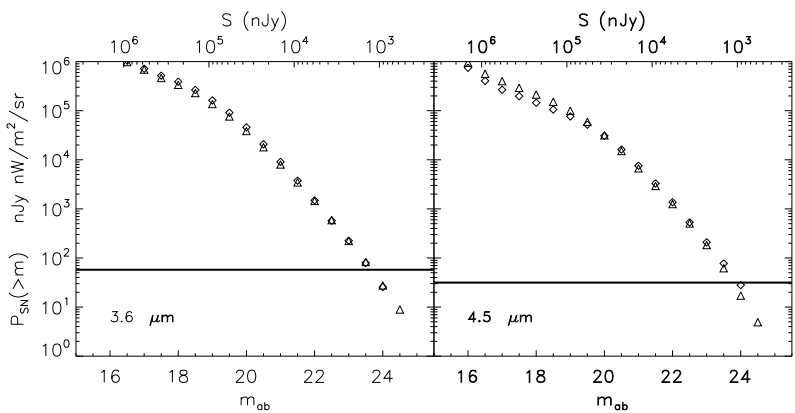} \vspace{0.5cm}
\caption[]{ \small{The cumulative shot noise of faint sources calculated from galaxy counts in the EGS (triangles) and UDS (diamonds) fields
are shown as a function of the faintest magnitude to which resolved sources are excluded. The horizontal line indicates the measured
shot noise as characterized by the power spectra at small spatial scales. The intersection implies that our analysis is excluding sources to
$m_{AB} \sim 24$, but this is a lower limit on the magnitude because the counts are becoming significantly incomplete at
$m_{AB} \gtrsim 22$. Reconstruction of  the shot-noise using a variety of multiband galaxy luminosity function data from Helgason et al. (2012) is shown with shaded areas and is immune to confusion; they find the AB magnitude limit around $\sim 25$ for these shot noise levels. }
} \label{fig:sn_counts}
\end{figure}

The derived galaxy counts, $dN(m)/dm$, were integrated from a lower magnitude limit to infinity to derive the shot noise
from the faint galaxies that may remain in the background, $P_{SN}(>m) = \int_{m_0}^\infty 10^{-0.8m} dN(m)/dm\ dm$.
Fig. \ref{fig:sn_counts} shows the shot noise levels from the measured counts as a function of the limiting magnitude, $m_0$.
The shot-noise levels shown in Fig. \ref{fig:sn_counts} are in excellent agreement with our earlier such determinations as shown in Fig. 1 of KAMM3. Note that the appropriate comparison there should be made to the shot noise levels reached in KAMM1 using comparable integrations; our analysis of GOODS data (KAMM2) with 25 hr/pixel total integrations reached shot noise levels a factor of $\sim 2-3$ lower.
Consistent with KAMM1-3 we conclude that the shot-noise level here implies that only sources at
$m_0 \gsim 24$
remain in the cleaned maps to contribute to the source-subtracted CIB clustering component at scales $\sim (20-2000)^{\prime\prime}$.
This shot noise calculation will suffer from incompleteness in the galaxy counts.
At faint magnitudes, $m_{\rm AB}\gsim 21$, the fields are confusion-limited (Fazio et al. 2004b) at the IRAC beam resolution of order $\sim (3-4)^{\prime\prime}$.  Correcting for incompleteness in this heavy confusion-limit leads to correction factors of $\gg 10$ and, as discussed in Fazio et al. (2004b), such corrections are suspect when corrections factors exceed $\sim 2$. Given this uncertainty,
the limiting magnitude is likely to increase to $m_0 \gsim 25.5-26$ after correcting for incompleteness as discussed in KAMM2. This is
supported by Helgason et al (2012), who reconstruct galaxy counts from multi-wavelength luminosity function data and derive similar
relations between the shot noise levels and the magnitude limits, $m_0$.


Are the sources contributing the clustered component just below the threshold of our source-removal, $\gsim m_0$, or is the clustered component produced by sources significantly fainter with $m\gg m_0$? The shot-noise can be expressed as $P_{\rm SN}\sim S F_{\rm CIB}(>m_0)$ where $S$ is the typical flux of the sources contributing to the measured fluctuations and producing the CIB of net flux $F_{\rm CIB}(>m_0)$ (KAMM3). Thus this can be addressed by comparison between the observed level of the large scale fluctuations, $\delta F_{\rm clus}$, and the
mean CIB level implied by the shot noise, $F_{\rm CIB}(>m_0) \sim P_{\rm SN}/S(m_0)$, where $S = 10^{-0.4(m_0-23.9)}\mu$Jy
is the typical flux of the population producing the large scale clustering component of the CIB. If both $P_{\rm SN}$ and
$\delta F_{\rm clus}$ are dominated by the brightest sources that were not masked or modeled, $m_0 \sim 24$, then
$\delta F_{\rm clus}/F_{\rm CIB}(>m_0) \sim 1$ at 3.6, 4.5 $\mu$m all the way to at least $\sim$degree scales. Thus, sources around  $m_0 \sim 24-25$ would need to be essentially completely
clustered in structures (and voids) with scales of $100'' - 1000''$. This strong large scale clustering is inconsistent with the
observed clustering of brighter sources and the predicted clustering of faint normal galaxy populations. Since the shot noise
level alone is consistent with normal galaxies at $m_0 \gtrsim 24$, we conclude that the large scale structure must arise from sources well below our threshold magnitude. 


\subsection{Contributions from remaining galaxy populations}

To date there are three estimates of the remaining known galaxy populations to the measured source-subtracted CIB fluctuations (KAMM1, Sullivan et al 2007, Helgason et al 2012). All of these estimates are consistent with each other and are discussed below.

KAMM1 have estimated the {\it upper} limit on the CIB fluctuations due to the remaining known galaxies as follows. They estimated the effective limiting magnitude of the remaining galaxies and used it to evaluate net CIB flux expected from the known galaxy populations using the counts of Fazio et al (2004). Then the upper limit on the residual CIB fluctuations were estimated by assuming that at all earlier epochs the remaining galaxies had the same clustering pattern as observed today on small scales with the 2-point correlation function $\xi(r)=(r/5.5h^{-1}{\rm Mpc})^{-1.7}$. This upper limit on the clustering component of the remaining CIB fluctuations is shown in Fig. \ref{fig:cib_final} after convolving with the IRAC beam as per Fig.1 of KAMM1.

In a different kind of analysis Sullivan et al (2007) have estimated the remaining CIB fluctuations at 3.6\um\ for the KAMM1 parameters by reconstructing the counts in shallow, but wide-field, IRAC measurements as well as deep GOODS observations, encompassing the angular scales of KAMM1. They measured the clustering of resolved sources out to $\sim 10^\prime$ down to AB magnitude of $\sim 24$ and then used a halo model (Cooray \& Sheth 2002) combined with conditional luminosity functions to estimate the CIB fluctuations at 3.6 \um\ from sources just below the threshold of the KAMM1 analysis. Their results are shown in Fig. \ref{fig:cib_final} and are taken from Fig. 8 of Sullivan et al (2007).

A different and substantially more extensive analysis was recently performed in Helgason et al (2012). They modelled the remaining CIB fluctuations using a massive compilation of 230 datasets of luminosity functions encompassing UV, optical and near-IR bands and spanning a wide range of redshifts. Using these data they were able to reconstruct empirically the evolution of the observed populations populating the redshift cone out to $z\sim 5$ at all wavelengths. The only uncertainties were due to the assumed extremes of the faint end of the luminosity function end, termed high-faint-end (HFE) and low-faint-end (LFE). The counts from these reconstructed populations were evaluated in the observer rest frames and were shown to be in excellent agreement with the measurements down to the faintest measured magnitude at all wavelengths from 0.45 to 4.5 \um. The shot-noise reconstructed from such data is immune to confusion. This, coupled with the concordance cosmological model ($\Lambda$CDM), allowed to robustly evaluate the source-subtracted CIB fluctuations at 3.6 and 4.5 \um\ due to the populations below the measured shot-noise levels for the HFE and LFE extremes. These are shown by shaded areas in Fig. \ref{fig:cib_final}.

Fig. \ref{fig:cib_final} shows the power spectrum expected for known galaxies from Sullivan et al. (2007) and Helgason et al. (2012) that are at fainter magnitudes than the shot noise constraint.
This contribution is significantly smaller than the measured level of the fluctuations at scales greater than $\sim 20^{\prime\prime}$ and also has a very different power spectrum. All of the estimates are mutually consistent and show that the existing galaxy populations remaining below the measured shot-noise levels cannot account for the clustering component of the CIB fluctuations measured by us.

\subsection{Unresolved very faint sources -- high-$z$ or low-$z$?}

The above discussion strongly suggests that the measured CIB fluctuations are produced by sources significantly fainter than the magnitude threshold of our source-removal procedure. In this case, it is likely that the measured shot-noise levels represent an {\it upper} limit on the shot-noise from these sources.
No such faint populations have been observed in the present Universe and, in addition, if they exist at low-$z$ these very faint cosmological sources would have to cluster very differently from the observed galaxy populations which would likely put them in conflict with the established $\Lambda$CDM cosmological model.

On the other hand, such faint populations are expected at high-$z$ epochs, but are
beyond the sensitivity and/or resolution of
current  telescopic studies. These populations would then be clustered according to the standard $\Lambda$CDM model except their clustering properties would be biased because they form in haloes identified with rare density peaks of the (overall) linear density field at these early times (Kashlinsky et al. 2004, Cooray et al. 2004, Fernandez et al. 2011).
To see if such populations can provide a reasonable fit to the measured large-scale clustering we model their power spectrum template as $P(q)=P_{\rm SN} + A_{5^\prime}^2 \; \frac{2\pi}{q_5^2}\;\frac{P_{\Lambda{\rm CDM}}(qd_A^{-1})}{P_{\Lambda{\rm CDM}}(q_5d_A^{-1})}$ where $P_{\Lambda{\rm CDM}}$ is a high-$z$ $\Lambda$CDM 3-dimensional template and $A_{5^\prime}$ is to-be-fitted amplitude at the fiducial scale of $2\pi/q_5\equiv5^\prime$ and $d_A$ is the comoving angular diameter distance to $z$. We emphasize the ``toy" nature of such a model: it assumes 1) that all the emission occurs over a narrow instant in time, and 2) the biasing is linear so that the 3-dimensional power spectrum of emitters is proportional to $P_{\Lambda{\rm CDM}}$. While the second assumption is fairly justifiable on the relevant linear scales, the first supposition is likely to be a significant over-simplification and in reality one needs to relate the projected angular power spectrum to the underlying 3-D one via the Limber equation after specifying the history and rate of flux production over a wide range of wavelengths.  Thus within the framework of this ``toy" model, the assumed templates are shown in Fig. \ref{fig:lcdm} and are practically independent of $z$ at the high redshifts. This is a consequence of the turn-over scale of the power spectra, which reflects the horizon at the matter-radiation equation in CDM-type models, subtending  essentially the same angular scale at $z\gg1$.

In detailed comparison of model and data, the effect of the masking on the power spectrum must be considered.
More specifically, the multiplication of the source subtracted images by a mask (values of 1 or 0), corresponds to a convolution
of the power spectrum by the Fourier transform of the mask. The effect would be redistributing some of the power from spatial scales
where it is stronger to those where it is weaker. This has little impact on a shot noise (and any other white) spectrum, but can modify the appearance
of a spectrum that rises sharply and has features of interest at large spatial scales. Thus the  $\Lambda$CDM-type
spectrum is affected more significantly than the spectrum of known galaxies. This is addressed in detail in Appendix B.
Because the spectrum of the fluctuations is measured from a masked map, the model has to be processed correspondingly.
More specifically, in Fourier space the Fourier transform of the model gets convolved with that of the mask and since the latter is known from our data, this transformation is unique. This processing is described in Appendix B, where Fig. \ref{fig:masking} shows the fluctuation spectra
after being processed through the mask.  Convolution with the mask does not affect the shot-noise component, and affects the spectrum of known galaxies only marginally. However, the $\Lambda$CDM-type spectrum is affected more significantly by the transfer of power due to mask if it arises at (high) redshifts such that the subtended comoving scales reflect its steeply changing part. (Some of the large-scale power will then be transferred to small scales $\lsim 10^{\prime\prime}$, but this is unobservable since it is overwhelmed by the much larger shot-noise power on these scales.)
The blue solid lines in Fig. \ref{fig:cib_final} show the fits to the data over scales $(100-1000)^{\prime\prime}$ with the effect of the masking included. The corresponding amplitudes of the model are $A_{5^\prime}$ = [0.07,0.05] nW/m$^2$/sr at [3,6, 4.5] $\mu$m respectively.
The sum of all three components (shot-noise plus known galaxies from Sullivan et al. (2007) plus the high-$z$ $\Lambda$CDM population) is shown in Fig. \ref{fig:cib_final} with the red solid line.

\section{Discussion}
\label{sec:conclusions}

The detected source-subtracted fluctuations appear of cosmological origin, but significantly exceed the signal from the observed known galaxy populations remaining in the data after the source-subtraction. The measured fluctuations are also in excellent agreement with our earlier measurements on the relevant scales and at the same level of shot-noise (KAMM1-2). Their spatial spectral distribution on sub-degree scales is also very different and is supportive of these fluctuations originating from sources located at early times and coincidental with the ``first stars era". These sources are individually inaccessible to current telescopic studies and this measurement allows for unique characterization of their properties and spatial distribution. An alternative to this high-$z$ interpretation would require the individual sources to be very faint as constrained by the observed shot-noise levels and the absence of correlations with the ACS sources (KAMM3), lie at low $z$ and have clustering properties significantly different from those of observed galaxy populations. No such a new population has been observed, or proposed on theoretical grounds, but if true this would represent an important discovery in its own right. The high-$z$ interpretation could be ruled out using optical and shorter wavelength IR imaging such as would be available from cross-correlating our maps with the HST WFC3 and ACS images. The presence of correlations between our IRAC source-subtracted maps and those in the optical would argue for the new low-$z$ populations.

Implications of the high-$z$ interpretation of the fluctuations have been discussed in KAMM3 and we briefly reiterate them here in light of the new measurements. The typical biasing expected for first haloes in the standard cosmology would lead to relative CIB fluctuations of at most $\sim 10\%$ on arcminute scales. Thus the measured fluctuations of less than $\sim 0.1$ nW/m$^2$/sr would require the net CIB flux at 3.5 and 4.5 \um\ of $\gsim (0.5-1)$ nW/m$^2$/sr. Such fluxes are well below the claimed mean CIB excess levels from the IRTS and DIRBE measurements around 3 \um\ (Dwek \& Arendt 1998, Matsumoto et al. 2005) whose theoretical implications have been addressed in Santos et al. (2002), Salvaterra \& Ferrara (2003) and Madau \& Silk (2005) among others. These are well within the current limits on mean CIB from the $\gamma$-ray absorption measurements in low-$z$ blazars (Dwek et al. 2005, Aharonian et al. 2005), although potentially this contribution may be detectable in the high-$z$  GRB spectra observed with {\it Fermi} LAT (Kashlinsky 2005b, Kashlinsky \& Band 2007, Gilmore 2011). These fluxes are also well within the uncertainties of CIB limits from HST deep observations of Thompson et al. (2007a). KAMM3 argue that these populations, having surface density of $\sim P_{\rm SN}/S^2$, must lie in the confusion noise of the present-day space-telescopes, and so care must be made when using filtering in assembling data from individual noisy exposures which may wash out these populations together with the unwanted noise (AKMM). Given the short cosmic time available at these epochs to produce the required CIB levels these sources must emit radiation much more efficiently with much lower $M/L$ than the present-day stellar populations (KAMM3) although there may be some admixture of lower-mass stars (Salvaterra et al. 2006). Even further, fluctuation measurements such as this do not directly require the sources of this emission to be exclusively stars and may contain additional contribution from the black-hole accretion emissions in the early Universe.

\acknowledgements
This material is based upon work supported by the NASA ADP and HST-C18 grants. We thank Kari Helgason for many useful discussions concerning the contributions from the observed galaxy populations to the measured signal.

{\it Facilities:} \facility{Spitzer (IRAC)}

\appendix
\section*{Appendix}
\setcounter{figure}{0}
\section{Comparison with previous results}

\begin{figure}[h!]
\renewcommand{\thefigure}{A-\arabic{figure}}
\centering \leavevmode \epsfxsize=1 \columnwidth
\epsfbox{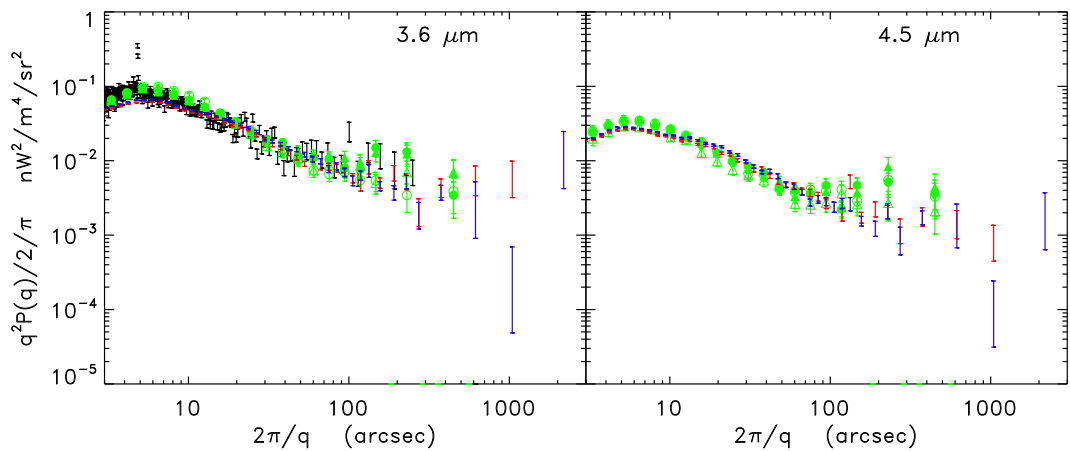} \vspace{0.5cm} \caption[]{\small{Comparison with KAMM1/KAMM2 measurements at the shot-noise levels reached here: red and blue error bars represent the UDS and EGS results as shown in Fig. \ref{fig:cib_fields}, black error bars correspond to the IOC/QSO 1700 field used in KAMM1, and green symbols show the four GOODS fields analyzed in KAMM2. Our earliest analysis in KAMM1 has not reached the shot-noise levels comparable to this study at 4.5 \um\ and, hence the data are not shown for clarity. However, Fig. 1 of KAMM2 demonstrates the consistency of that measurement with the four additional GOODS field shown here at the shot-noise levels comparable to this study.} } \label{fig:comparison}
\end{figure}

Unlike Galactic or Solar System foregrounds, diffuse light fluctuations due to sources at cosmological distances should be isotropically distributed on the sky. The signal should then be similar at all fields provided the foreground sources are removed to approximately the same shot-noise levels. Fig. \ref{fig:comparison} compares all the measurements obtained by us at comparable shot-noise levels (to within $\sim 30\%$)\footnote{Table 1 of KAMM2 gives the shot-noise levels and Fig. 1, left of KAMM4 shows how $P_{\rm SN}$ decrease with Model iteration. In these earlier analyses we used coarser grid of saved iterations than here, so we chose the shot-noise closest to the one reached in this study: $P_{\rm SN} \sim (4.8, 2.2) \times 10^{-11}$ nW$^2$/m$^4$/sr at (3.6, 4.5) \um.}. For historical reasons, and because of its shallower exposure, the field studied in KAMM1 at 4.5 \um\ had populations removed only to much larger shot-noise level of $P_{\rm SN} \simeq 6\times 10^{-11}$ nW$^2$/m$^4$/sr and hence is not shown in the right panel of the figure; however, its consistency with the KAMM2 measurements in the four GOODS fields at the same shot-noise level is shown in Fig. 1 of KAMM2.

\section{Correcting model templates for masking}
\label{sec:masking}

The power spectra are evaluated from the masked images. Although in our case the fraction of pixels lost to the mask is small ($\simeq 26\%$), it is nonetheless important to evaluate its effects when comparing to predictions of a given input model.

\begin{figure}[h!]
\renewcommand{\thefigure}{A-\arabic{figure}}
\centering \leavevmode \epsfxsize=0.65 \columnwidth
\epsfbox{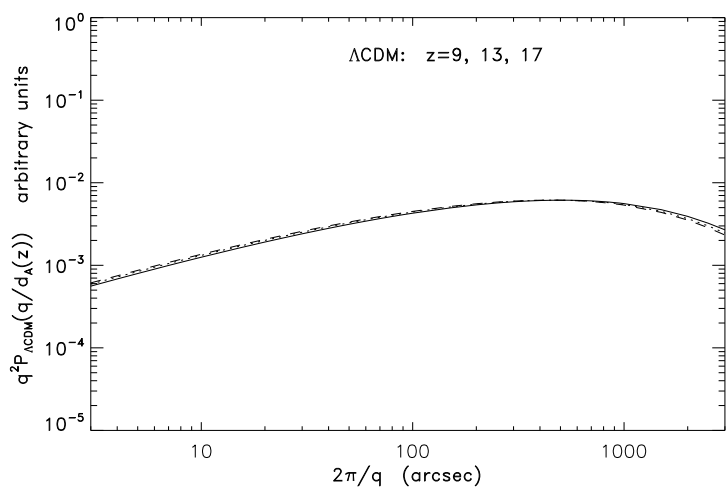} \caption[]{\small{High-$z$ shape of the concordance $\Lambda$CDM power spectrum projected to $z=9, 13, 17$.} } \label{fig:lcdm}
\end{figure}
There are three potential contributions to the measured fluctuation spectrum coming from 1) shot-noise from remaining sources, 2) clustering of known galaxies, and 3) from high-$z$ sources modelled with $\Lambda$CDM spectrum. The galaxy clustering contribution template we adopt per Sullivan et al. (2007) and Helgason et al. (2011). The template of the putative high-$z$ component of sources with the $\Lambda$CDM is shown in Fig. \ref{fig:lcdm}.

All of these assumed components are measured from a map which is masked after clipping the resolved sources. Because the mask is known, this effect can and should be corrected for leading to unique transformation of the assumed template of the power spectrum.

\begin{figure}[h!]
\renewcommand{\thefigure}{A-\arabic{figure}}
\centering \leavevmode \epsfxsize=0.8 \columnwidth
\epsfbox{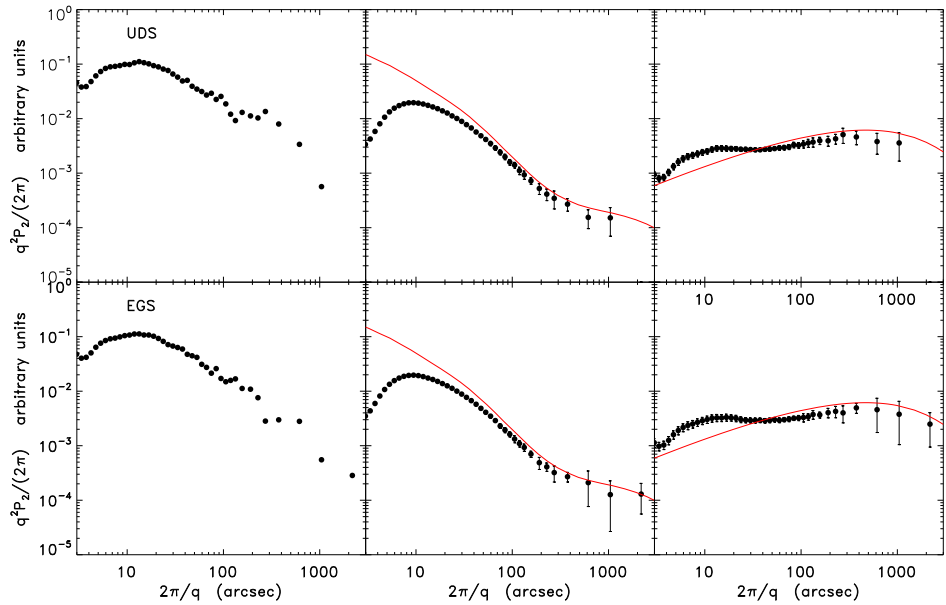} \vspace{0.5cm} \caption[]{\small{Masking effects on fluctuations with different power spectra. The results are based on 100 realizations of CIB with a given power spectrum and field parameters. Filled dots show the mean of the realizations and error bars correspond to one standard deviation. Left: Power spectrum of the mask. Middle: Known galaxies with the spatial power spectrum from Sullivan et al. (2007). The input template {\it without} the mask convolution is shown with a red line. Right: high-$z$ $\Lambda$CDM. The input template {\it without} the mask convolution is shown with a red line.  (The convolution with the instrument beam is omitted in all cases.)} } \label{fig:masking}
\end{figure}

The measured fluctuations are multiplied by the mask template in real angular space. This is equivalent to convolution in Fourier space. Thus the shot-noise will remain flat (except it is now multiplied by the mask window instead of the beam), while the other components with $P\neq$const will be convolved in a complicated way with the mask. To understand the effects of the mask and to obtain the templates after mask-processing we have conducted simulations generating many realizations of a diffuse field of appropriate geometric configuration with a given power spectrum, masking it and computing the resultant power spectrum and its distribution.
Fig. \ref{fig:masking} shows the effects of the EGS and UDS masks on the simulated CIB spectra of the various components.

\begin{deluxetable}{ccccccc}
\tablecaption{Analyzed SEDS Fields}
\tablehead{
\colhead{Region} & \colhead{$\alpha,\delta$} & \colhead{$l_{\rm Gal},b_{\rm Gal}$} &
\colhead{$\lambda_{\rm Ecl},\beta_{\rm Ecl}$} & \colhead{Size} &
\colhead{$\langle t_{\rm obs} \rangle$} & \colhead{$f_{\rm sky}$} \\
\colhead{}& \colhead{(deg)}  & \colhead{(deg)} & \colhead{(deg)} & \colhead{} &
\colhead{(hr)} & \colhead{}
}
\startdata
UDS &  34.50, -5.17 & 169.98, -59.88 &  30.41, -17.88 & $21^\prime\times21^\prime$ & 13.6 & 0.730\\
EGS & 214.91, 52.43 &  95.95,  59.81 & 180.56,  60.00 &  $8^\prime\times62^\prime$ & 12.5 & 0.725\\
\enddata
\label{tab:table1}
\tablecomments{The fields are located at moderate to high Galactic latitudes to minimize the number of foregrounds stars
and the brightness of the emission from interstellar medium (cirrus). These fields also lie at relatively high ecliptic latitudes,
which helps minimize the brightness and temporal change in the zodiacal light from interplanetary dust. The observations of each field are
carried out at three different epochs, spaced 6 months apart.}
\end{deluxetable}

\end{document}